\documentstyle[12pt]{article}
\begin{document}
%--------------------------------------
\newcommand{\slh}{\hspace{-.8em}/}
\newcommand{\Tr}{\mbox{Tr}}
\newcommand{\Real}{\mbox{Re}}
\newcommand{\Li}[1]{\mbox{Li}_{#1}}
\newcommand{\Had}{\mbox{Had}}
\newcommand{\Lep}{\mbox{Lep}}
\newcommand{\eins}{1\hspace{-.35em}1}
\renewcommand{\textfraction}{0}
\renewcommand{\topfraction}{1}
\renewcommand{\bottomfraction}{1}
\renewcommand{\floatpagefraction}{0}
\pagestyle{plain}
\sloppy
%-----------------------------
\thispagestyle{empty}
\vspace*{-2mm}
\thispagestyle{empty}
\noindent
\hfill KA--TP--3--1995\\
\mbox{}
\hfill  March 1995   \\
\vspace{0.5cm}
\begin{center}
  \begin{Large}
  \begin{bf}
Supersymmetric Box Contributions to \\
$ e^+ e^- \rightarrow Z^0 h^0 $ and $ e^+ e^- \rightarrow A^0 h^0 $.\\
  \end{bf}
  \end{Large}
  \vspace{0.8cm}
  \begin{large}
  Volker DRIESEN and Wolfgang HOLLIK\\[5mm]
    Institut f\"ur Theoretische Physik\\
    Universit\"at Karlsruhe\\
    Kaiserstr. 12,    Postfach 6980\\[2mm]
    D-76128 Karlsruhe, Germany\\
  \end{large}
  \vspace{2.5cm}
  {\bf Abstract}
\end{center}
\begin{quotation}
\begin{center}
\parbox{12.cm}{
Within the MSSM we calculate the electroweak 1-loop box
contributions to the processes
$ e^+ e^- \rightarrow Z^0 h^0 $ and $ e^+ e^- \rightarrow A^0 h^0 $.
We present detailed results for c. m. energies $\sqrt{s}=200$ GeV
and $\sqrt{s}=500$ GeV as well as for $\tan\beta=2$ and $\tan\beta=50$.
The box contributions to the process $ e^+ e^- \rightarrow Z^0 h^0 $
are, depending on $\sqrt{s}$ and $\tan\beta$, of the order -2 to -20\% and
to $ e^+ e^- \rightarrow A^0 h^0 $ of 2 to 10\%.
}
\end{center}
\end{quotation}
\newpage
%------------------------------------------
\setcounter{page}{1}
\section{Introduction}

Experiments at the forthcoming upgrade of the LEP collider to LEP2 will
continue the search for Higgs bosons in the mass range up to 100-110 GeV
\cite{R1}, depending on the energy and luminosity at this machine.
Besides the search for the Standard Model Higgs boson,
numerous studies are devoted
to the Higgs bosons of the minimal supersymmetric standard model (MSSM)
\cite{R2}, the most predictive framework beyond the Standard Model,
where at least one light neutral scalar with mass $\stackrel{<}{\sim}$
130 GeV is expected. This mass range can be fully explored at new $e^+ e^-$
colliders with energies up to 500 GeV \cite{R2,R7}. Detailed studies of
Higgs production and decay processes are required to detect the possible
signals of Higgs bosons and to distinguish as far as possible the origin
of a produced scalar particle. The most promising process to search for the
standard Higgs boson is the bremsstrahlung type reaction
$e^+ e^- \rightarrow Z H$
at energies $> M_H + M_Z$. For Higgs bosons of the MSSM the basic
production channels are
\begin{displaymath}
\mbox{}\hspace{6.cm} e^+ e^- \rightarrow Z h^0 \hspace{8.cm} (Z)
\end{displaymath}
and
\begin{displaymath}
\mbox{}\hspace{6.cm} e^+ e^- \rightarrow A^0 h^0 \hspace{7.9cm} (A)
\end{displaymath}
where $h^0$ is the lightest CP even and $A^0$ the CP odd neutral particle
of the MSSM Higgs spectrum.

For an accurate discussion of the production
cross sections one has to include radiative corrections. For the standard
Higgs complete 1-loop calculations for $e^+ e^- \rightarrow Z H$ are
available \cite{R3}.
For the corresponding processes in the MSSM no complete 1-loop calculation
has been performed so far.
Besides the approximate methods of the effective potential approach \cite{R4}
and the renormalization group treatment \cite{R5} it is of particular
interest to also have a complete diagrammatic calculation:
\begin{itemize}
\item[-]
it allows for all virtual particle effects within the MSSM without
restrictions to masses and mixings,
\item[-]
it takes into account all momentum dependent effects in 2- and 3- point
functions,
\item[-]
it provides a reference frame for checking the quality of simpler compact
approximations which might be useful for practical applications.
\end{itemize}
The complete set of 1-loop diagrams contributing to the MSSM neutral
Higgs production processes is very extensive, and until now only one
calculation has been carried out \cite{R6} based on the following building
blocks: the set of self energy corrections to the vector boson and
Higgs propagators and wave functions, the vertex corrections to the
$e^+ e^- Z$, $Z Z h$, $Z A h$ vertices, and the corresponding counter
terms in the on--shell renormalization scheme. The box diagram
contributions have not yet been considered so far, and they are also
not part of the phenomenological studies for LEP2 and higher energy
colliders \cite{R7,R8}.

In this paper we provide the box contributions to both
$e^+ e^- \rightarrow Z h$ and $e^+ e^- \rightarrow A h$. Box diagrams
are required for theoretical reasons to render the amplitudes gauge
independent, and for practical reasons since they are in general of
non-negligible size. The computation in \cite{R6} on the basis of 2- and
3-point functions was performed in the \mbox{}`t Hooft-Feynman gauge with
$\xi=1$. In this specific gauge, the box contributions are finite and
independent of the details of the renormalization. Hence, we proceed to
derive the box contribution in the $R_{\xi=1}$ gauge as well, thus
providing the missing ingredients as an independent, finite and scheme
independent block for completing the amplitudes and cross sections.

%------------------------------------------
\section{Cross Sections}

The Born diagrams contributing to process (Z) and (A) are displayed in
Fig. 1 together with the momentum assignment. $p_3$ always denotes the
outgoing $h^0$ momentum,
whereas $p_4$ is the momentum of the outgoing $Z$ or $A^0$,
respectively. $\varepsilon^\mu_Z$ denotes the polarization vector of the
$Z$. Furthermore, we use the invariant variables
\begin{displaymath}
s=(p_1+p_2)^2, \hspace{1.cm}
t=(p_1-p_3)^2, \hspace{1.cm}
u=(p_1-p_4)^2
\end{displaymath}
and the scattering angle $\theta = \;<{\hspace{-.8em})}(e^-,h^0)$ in the
c. m. system. The masses of $e^+$ and $e^-$ have been neglected for the
entire calculation.

Including the box diagrams (Fig. 2 and 3) one has to take into account
their interference with the corresponding Born amplitudes. The differential
cross section in the c. m. system can be written in the following way:
\begin{eqnarray}
 \frac{ \mbox{d} \sigma_{Z}}{\mbox{d} \cos\theta} & = &
 \frac{\sqrt{\lambda(s,M_Z^2,M_h^2)}}{32 \pi s^2}
  \bigg(T^0_Z + \sum_n T_Z^n \bigg) \label{E1}\\
 \frac{ \mbox{d} \sigma_{A}}{\mbox{d} \cos\theta} & = &
 \frac{\sqrt{\lambda(s,M_A^2,M_h^2)}}{32 \pi s^2}
  \bigg(T^0_A + \sum_l T_A^l \bigg) \label{E2}
\end{eqnarray}
with $\lambda(x,y,z) = x^2+y^2+z^2-2(xy+xz+yz)$. The index $n$ counts the
diagrams of Fig. 2, and $l$ those of Fig. 3. Thereby it is understood that
the summation is performed also over the various
$\widetilde{\chi}^0$ and $\widetilde{\chi}^+$ states as well as over
$\widetilde{e_L}$ and $\widetilde{e_R}$.

$T_{Z,A}^0$ denote the lowest
order contributions following from the Born amplitudes
\begin{equation}
 {\cal M}_{Z,A}^0 \;=\; \overline{v}(p_2) \;{\cal O}_{Z,A}^0\; u(p_1)
\end{equation}
with
\begin{equation}
{\cal O}_Z^0 \;=\;
 i \frac{g^2 M_Z^2 \sin(\beta-\alpha)}{c_w^2 (s-M_z^2)}
 \;\gamma_{\mu}\;
    \Big[(g_V^e+g_A^e) {\cal P}_L+(g_V^e-g_A^e) {\cal P}_R\Big]
    \:\varepsilon^{\mu}_Z (\sigma)
\end{equation}
for process (Z), and
\begin{equation}
{\cal O}_A^0 \;=\;
 - \frac{g^2 \cos(\beta-\alpha)}{2\: c_w^2 (s-M_z^2)}
 \;\gamma_{\mu}\;
   \Big[(g_V^e+g_A^e) {\cal P}_L+(g_V^e-g_A^e) {\cal P}_R \Big]
   \: (p_3 - p_4)^{\mu}
\end{equation}
for process (A), where
\begin{equation}
g_V^e \;=\; -\frac{1}{4}+s_w^2\;,\hspace{1.cm}
g_A^e \;=\; -\frac{1}{4}
\end{equation}
are the standard NC couplings, $g=e/s_w$, and $\alpha$, $\beta$ the
mixing angles of the Higgs sector.
\\

Averaging over the initial helicities for unpolarized $e^\pm$
and summing over the $Z$ polarizations $\sigma$ in case of (Z)
yield the following expressions
\begin{eqnarray}
T_{Z}^0 &=& \frac{1}{4} \sum_{\sigma} \Tr  \left( p_2\slh \;\;
 {\cal O}_{Z}^0\; p_1\slh\;\; \overline{\cal O}_{Z}^0 \right)
 \label{E71}\\
T_{A}^0 &=& \frac{1}{4} \Tr  \left( p_2\slh \;\;
 {\cal O}_{A}^0\; p_1\slh\;\; \overline{\cal O}_{A}^0 \right)
\label{E72}
\end{eqnarray}
which result in the explicit forms of the Born cross sections:
\begin{eqnarray}
 \frac{ \mbox{d} \sigma^0_{Z}}{\mbox{d} \cos\theta} & = &
 \frac{1}{128 \pi s^2} \frac{g^4}{c_w^4} \sin^2(\beta-\alpha)
 \frac{\sqrt{\lambda(s,M_Z^2,M_h^2)}}{(s-M_Z^2)^2}
 \Big({g_V^e}^2+{g_A^e}^2\Big) \:2\: \Big(t\:u + 2 s M_Z^2 -M_Z^2 M_h^2\Big)
 \label{E9} \\[3mm]
 \frac{ \mbox{d} \sigma^0_{A}}{\mbox{d} \cos\theta} & = &
 \frac{1}{128 \pi s^2} \frac{g^4}{c_w^4} \cos^2(\beta-\alpha)
 \frac{\sqrt{\lambda(s,M_A^2,M_h^2)}}{(s-M_Z^2)^2}
 \Big({g_V^e}^2+{g_A^e}^2\Big) \:2\: \Big(t\:u  - M_A^2 M_h^2\Big)
 \label{E10}
\end{eqnarray}
The interference terms in (\ref{E1}), (\ref{E2}) can be expressed in a way
analogous to (\ref{E71},\ref{E72}):
\begin{eqnarray}
T_{Z}^n & = & \frac{1}{2}\; \sum_{\sigma} \Real \left[
          \Tr \left( p_2\slh\;\; {\cal O}_{Z}^n\; p_1\slh \;\;
                   \overline{\cal O}_{Z}^0 \right) \right] \\
T_{A}^l & = & \frac{1}{2}\; \Real \left[
          \Tr \left( p_2\slh\;\; {\cal O}_{A}^l\; p_1\slh \;\;
                   \overline{\cal O}_{A}^0 \right) \right]
\end{eqnarray}
with the operators ${\cal O}_Z^n$ corresponding to the diagrams of Fig. 2
and  ${\cal O}_A^l$ to those of Fig. 3.

For the discussion of the various contributions it is useful to
distinguish between the genuine SUSY diagrams with virtual selectrons (se)
and sneutrinos (snu) and the residual diagrams of the standard model with
2 Higgs doublets (2hdm):\\
Figure 2:
\begin{eqnarray}
 {\cal O}_Z^{\mbox{\scriptsize se}}
  &=& \mbox{diagrams}(1,2,3)+(h\leftrightarrow Z) \nonumber\\
 {\cal O}_Z^{\mbox{\scriptsize snu}}
  &=& \mbox{diagrams}(4,5,6)+(h\leftrightarrow Z) \label{OPZ}\\
 {\cal O}_Z^{\mbox{\scriptsize 2hdm}}
  &=& \mbox{diagrams}(7,8)+(h\leftrightarrow Z)
                       +\mbox{diagrams}(9,10) \nonumber
\end{eqnarray}
Figure 3:
\begin{eqnarray}
{\cal O}_A^{\mbox{\scriptsize se}}
  &=& \mbox{diagram }2+(h\leftrightarrow A)+
                   \mbox{diagram }4\nonumber\\
{\cal O}_A^{\mbox{\scriptsize snu}}
  &=& \mbox{diagram }3+(h\leftrightarrow A)+\mbox{diagram }5 \label{OPA}\\
{\cal O}_A^{\mbox{\scriptsize 2hdm}}
  &=& \mbox{diagram }1+(h\leftrightarrow A)\nonumber
\end{eqnarray}
including summation over the $\widetilde{\chi}^0$,
$\widetilde{\chi}^+$, and $\widetilde{e}_{L,R}$ states.

Our conventions for the Feynman rules are adopted from ref. \cite{HHG}
and for fermion number violating processes from ref. \cite{DennerRules}.
The calculations were done using the algebraic program FORM \cite{FORM}
and the package FF \cite{FF} to evaluate the tensor integrals.

According to the decomposition (\ref{OPZ},\ref{OPA})
we can write for the cross sections:
\begin{equation}
\frac{\mbox{d} \sigma_{Z,A}}{\mbox{d}\cos\theta} \;=\;
\frac{\mbox{d} \sigma_{Z,A}^0}{\mbox{d}\cos\theta}
\Big[ 1+\delta^{\mbox{\scriptsize box}}_{Z,A} \Big]
\end{equation}
with the relative box contribution
\begin{equation}
\delta^{\mbox{\scriptsize box}}_{Z,A} \;=\;
\delta^{\mbox{\scriptsize 2hdm}}_{Z,A} +
\delta^{\mbox{\scriptsize se}}_{Z,A} +
\delta^{\mbox{\scriptsize snu}}_{Z,A} \label{E14}.
\end{equation}
In a similar way we express the integrated cross sections as
\begin{equation}
 \sigma_{Z,A} \;=\;
 \sigma_{Z,A}^0
\Big[ 1+\Delta^{\mbox{\scriptsize box}}_{Z,A} \Big]
\end{equation}
with
\begin{equation}
\Delta^{\mbox{\scriptsize box}}_{Z,A} \;=\;
\Delta^{\mbox{\scriptsize 2hdm}}_{Z,A} +
\Delta^{\mbox{\scriptsize se}}_{Z,A} +
\Delta^{\mbox{\scriptsize snu}}_{Z,A}
\end{equation}
where $\sigma^{0}_{Z,A}$ denote the Born terms following from
(\ref{E9}) and (\ref{E10}) by integration over $\cos\theta$.

%------------------------------------------
%
\section{Results and discussion}

In this section we present and discuss the numerical influence of the
box diagrams on the production cross sections. As concrete examples we
have chosen two sets of Higgs masses where both $Zh^0$ and $A^0h^0$ can
be produced at LEP2 energies. Taking $M_A=90$ GeV, one has $M_h=40$ GeV
for $\tan \beta=2$ and $M_h=89$ GeV for $\tan \beta=50$. These mass
values correspond to the lowest order MSSM Higgs spectrum. At this
stage we have not yet included higher order effects since that would
include a bunch of additional parameters like $m_t$, scalar top masses
and mixing parameters which are not directly related to the specific
problems addressed in this note. The input required for calculating
the box contributions is $M_h$ and the mixing angle $\alpha$, besides
$M_A$, $\tan\beta$, the slepton and chargino/neutralino parameters. In
a full calculation $M_h$ and $\alpha$ will be derived from the dressed
$( h^0,H^0)$ propagator system and thus are provided automatically
 in a comprehensive
one-loop program. The simplified approach corresponds roughly to a
situation where the scalar top system has masses around $m_t$, i.e.
$m_{\widetilde{t_1}}\cdot m_{\widetilde{t_2}} \cong m_t^2$, when no
large corrections to neutral Higgs masses and couplings occur.
\\

\noindent
We discuss the two processes (Z) and (A) seperately:\\[2mm]

\noindent
{\bf\large $e^+ e^- \rightarrow Z h^0$: }\\

\noindent
In order to exhibit the size of the box contribution we show in Fig. 4
the relative correction $\delta^{\mbox{\scriptsize box}}_{Z}$, eq.
(\ref{E14}), in the differential cross section as function of the
scattering
angle $\theta$ for two energies, $\sqrt{s}=200$ GeV and $500$ GeV. For
the scalar leptons a universal mass $m_{\widetilde{l}}=100$ GeV and no
mixing is assumed, and the $\widetilde{\chi}^0,\widetilde{\chi}^+$
system is fixed by choosing $M=\mu=150$ GeV. This corresponds to a mass
of $76$ GeV for the lightest $\widetilde{\chi}^+$ and $44$ GeV to the
lightest $\widetilde{\chi}^0$. The figure shows that the contribution
reaches already $-5\%$ at $200$ GeV for large $\tan\beta$, and increases
remarkably to $-(20\div 25)$\% at $500$ GeV. It can be seen that the
dominating piece is from the gauge and Higgs sector (2hdm), whereas the
genuine SUSY boxes are smaller than 5\%, also when we decrease the slepton
masses to lower values close to the present experimental limit.
This dominance of the gauge-Higgs part
persists also at higher energies, as displayed in Figure 5
showing the integrated cross section as function of the c. m. energy.
The dependence on the SUSY breaking parameters $M,\mu$ is visualized in
Fig. 6, again for the integrated cross section: for increasing $M$ and
$\mu$ the genuine SUSY boxes become very small, as expected from
decoupling properties, and we are left with the gauge-Higgs subclass
of diagrams. As a consequence, the box contribution to
$e^+ e^- \rightarrow Z h$ is not very sensitive to the details of the
SUSY sector, and is qualitatively very similar to the Standard Model
situation \cite{R3}.\\[2mm]

\noindent
{\bf\large $e^+ e^- \rightarrow A^0 h^0$: }\\

\noindent
In Fig. 7 we show
the relative correction $\delta^{\mbox{\scriptsize box}}_{A}$
(eq. \ref{E14}), with the same parameter choices as in Fig. 4.
The contribution of the box diagrams to the process (A) are somewhat
smaller than to the process (Z): at $\sqrt{s}=200$ GeV the relative
correction is about 2\% and reaches nearly 10\% at 500 GeV. Again the
contributions of the  gauge and Higgs sector dominate. The effect of
diagrams containing genuine SUSY particles is less than $0.5\div 1$\%
for $m_{\widetilde{l}}=100$ GeV and slightly larger for
$m_{\widetilde{l}}=50$ GeV.
The dependence of the integrated cross section on the c. m. energy
$\sqrt{s}$ is plotted in Fig. 8. As can be seen, the effects of genuine
SUSY diagrams decrease with $\sqrt{s}$. We also show the
dependence on $M$ and $\mu$ in Fig. 9: the behaviour is similar to
the case of the $Zh^0$ production, i.e. the SUSY boxes become very small
with increasing SUSY breaking parameters.\\[2mm]

In summary, the box diagrams yield a sizable contribution to the
differential and integrated Higgs production cross sections, in particular
in the energy domain of a next linear $e^+e^-$ collider.
\newpage
\clearpage

% -----------------------------------------
\def\npb#1#2#3{{\it Nucl. Phys. }{\bf B #1} (#2) #3}
\def\plb#1#2#3{{\it Phys. Lett. }{\bf #1 B} (#2) #3}
\def\prd#1#2#3{{\it Phys. Rev. }{\bf D #1} (#2) #3}
\def\prl#1#2#3{{\it Phys. Rev. Lett. }{\bf #1} (#2) #3}
\def\prc#1#2#3{{\it Phys. Reports }{\bf C #1} (#2) #3}
\def\pr#1#2#3{{\it Phys. Reports }{\bf #1} (#2) #3}
\def\zpc#1#2#3{{\it Z. Phys. }{\bf C #1} (#2) #3}
\def\nca#1#2#3{{\it Nouvo~Cim.~}{\bf #1A} (#2) #3}
%

% -----------------------------------------
\newpage
{\bf \Huge Figures}
\begin{itemize}
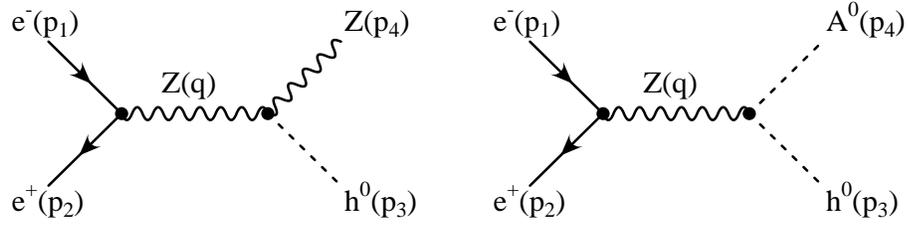
\item{\bf Fig.1}
   Born diagrams contributing to $ e^+ e^- \rightarrow Z^0 h^0 $
   and $ e^+ e^- \rightarrow A^0 h^0 $.
\item{\bf Fig.2}
   Box diagrams contributing to $ e^+ e^- \rightarrow Z^0 h^0 $.
\item{\bf Fig.3}
   Box diagrams contributing to $ e^+ e^- \rightarrow A^0 h^0 $.
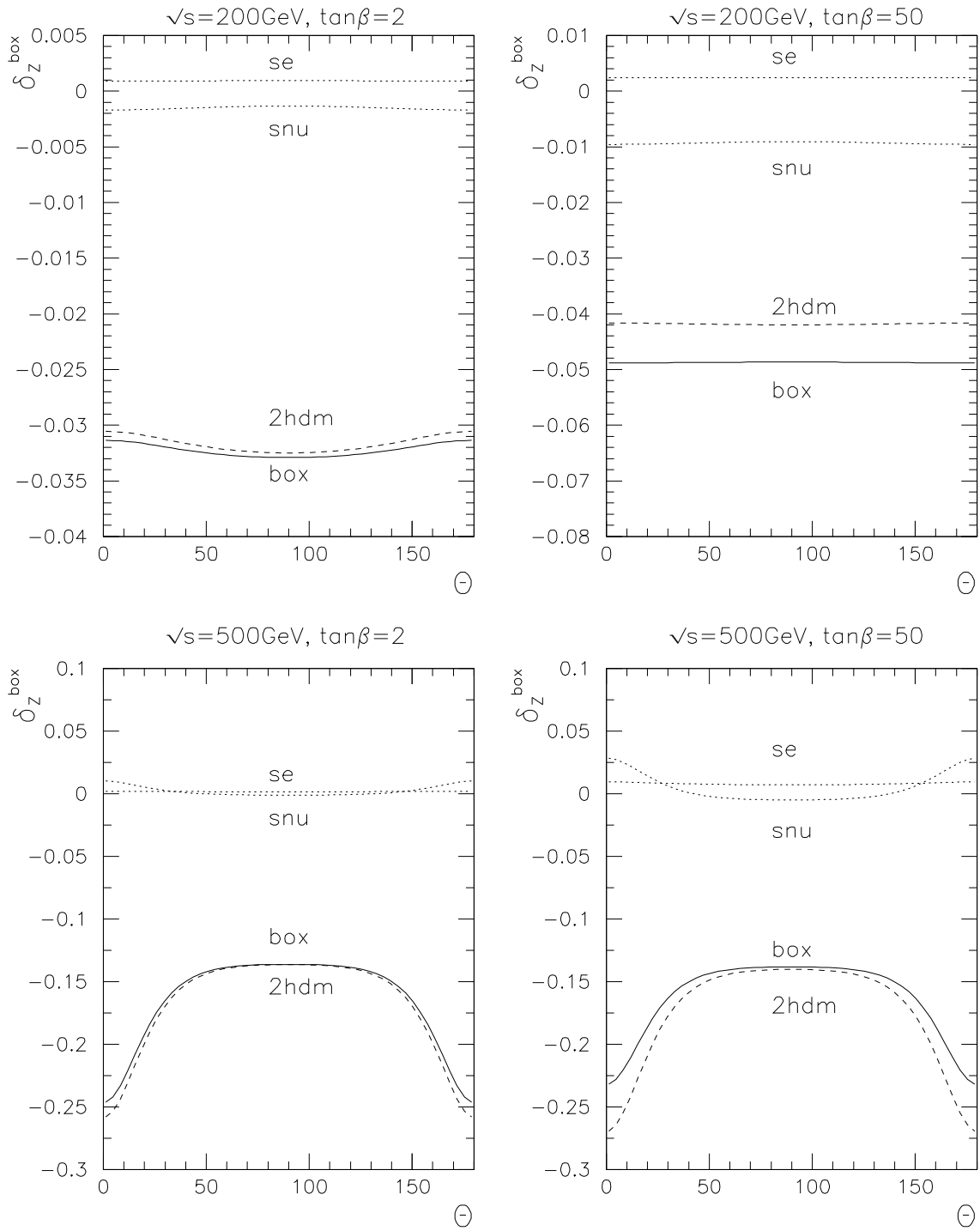
\item{\bf Fig.4}  Relative box contribution $\delta^{box}_Z$
  for  $ e^+ e^- \rightarrow Z^0 h^0 $
  at $\sqrt{s}=200$ GeV, 500 GeV and $\tan\beta=2$, 50.
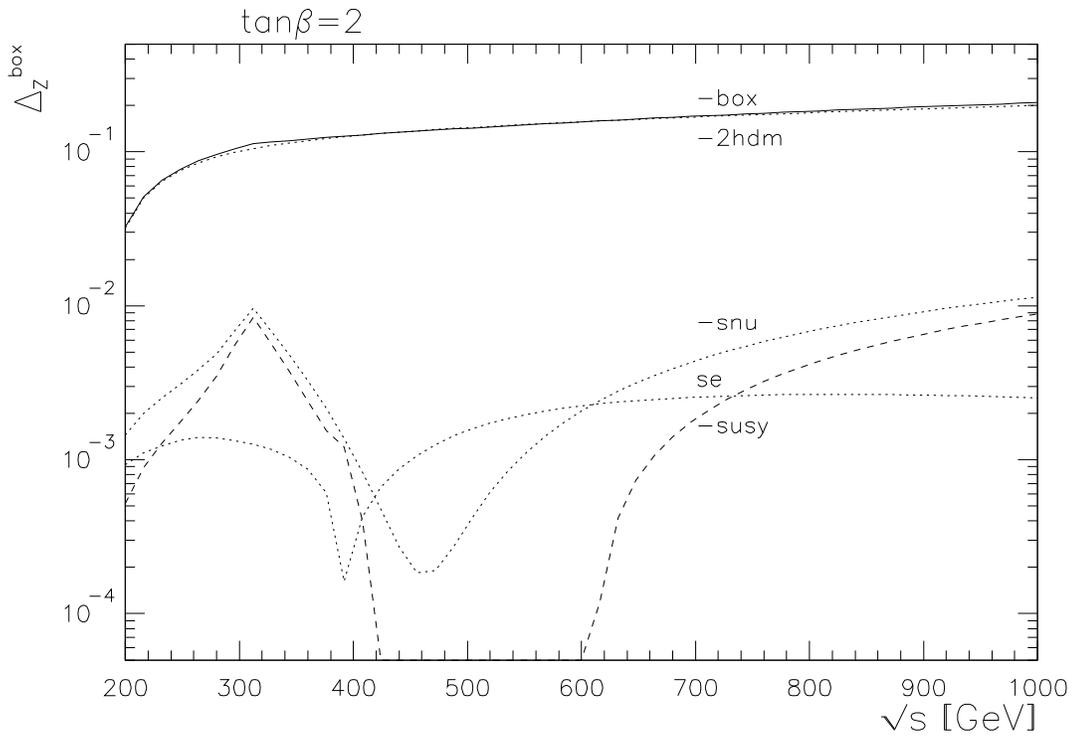
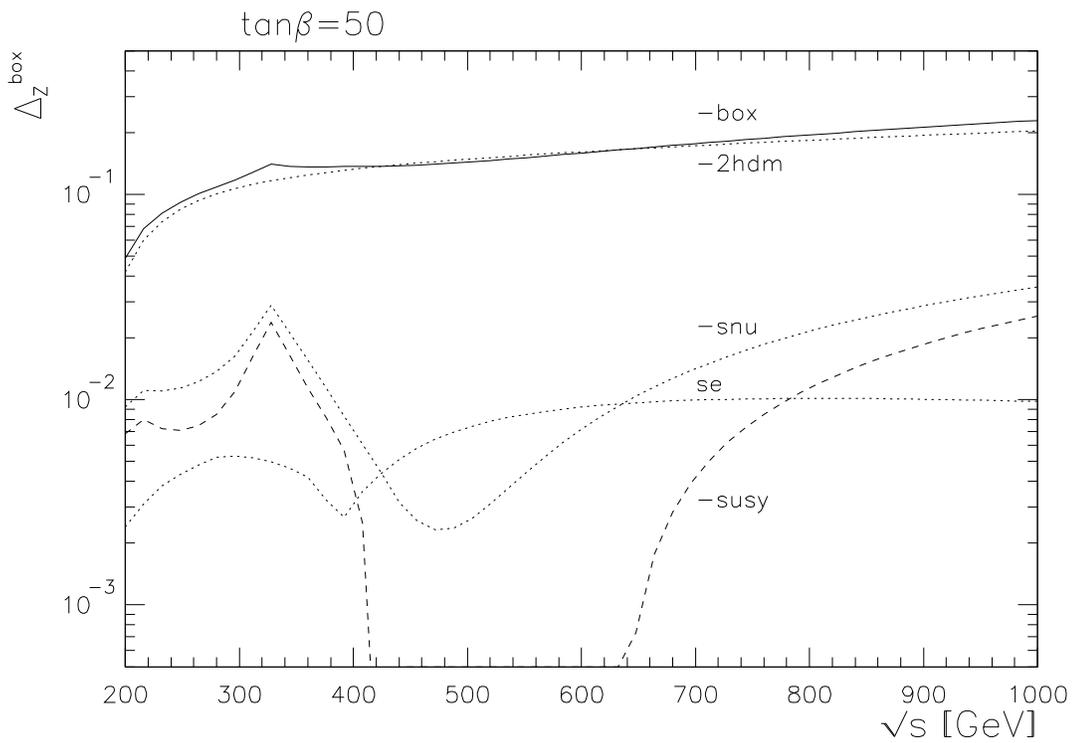
\item{\bf Fig.5}  Relative box contribution $\Delta^{box}_Z$
  for $M=\mu=150$ GeV.\\
  If curves are marked by `-se' etc.
  $-\Delta^{\mbox{\scriptsize se}}_{Z,A}$ is plotted.
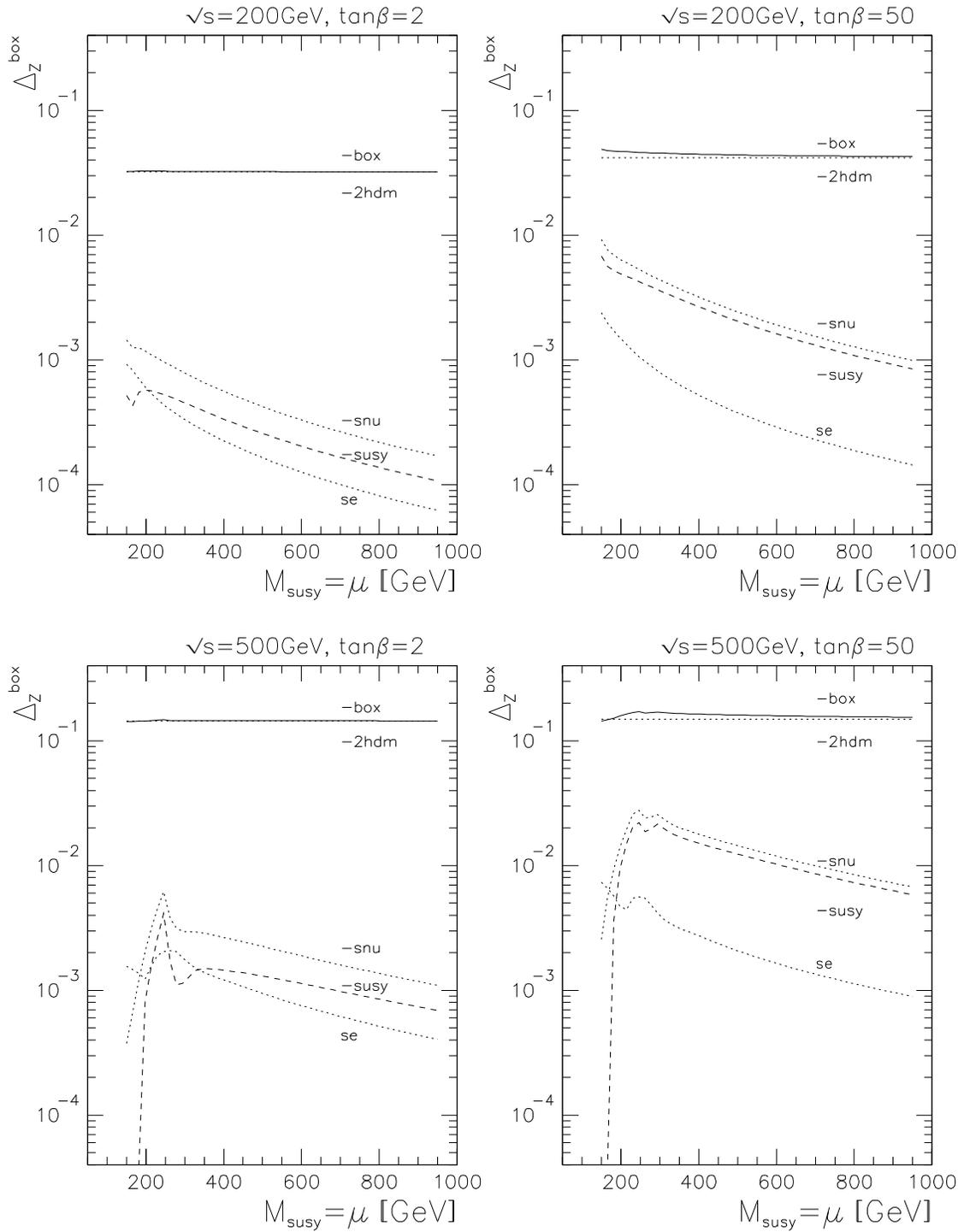
\item{\bf Fig.6}   $\Delta^{box}_Z$
  at $\sqrt{s}=200$ GeV, 500 GeV and $\tan\beta=2$, 50.
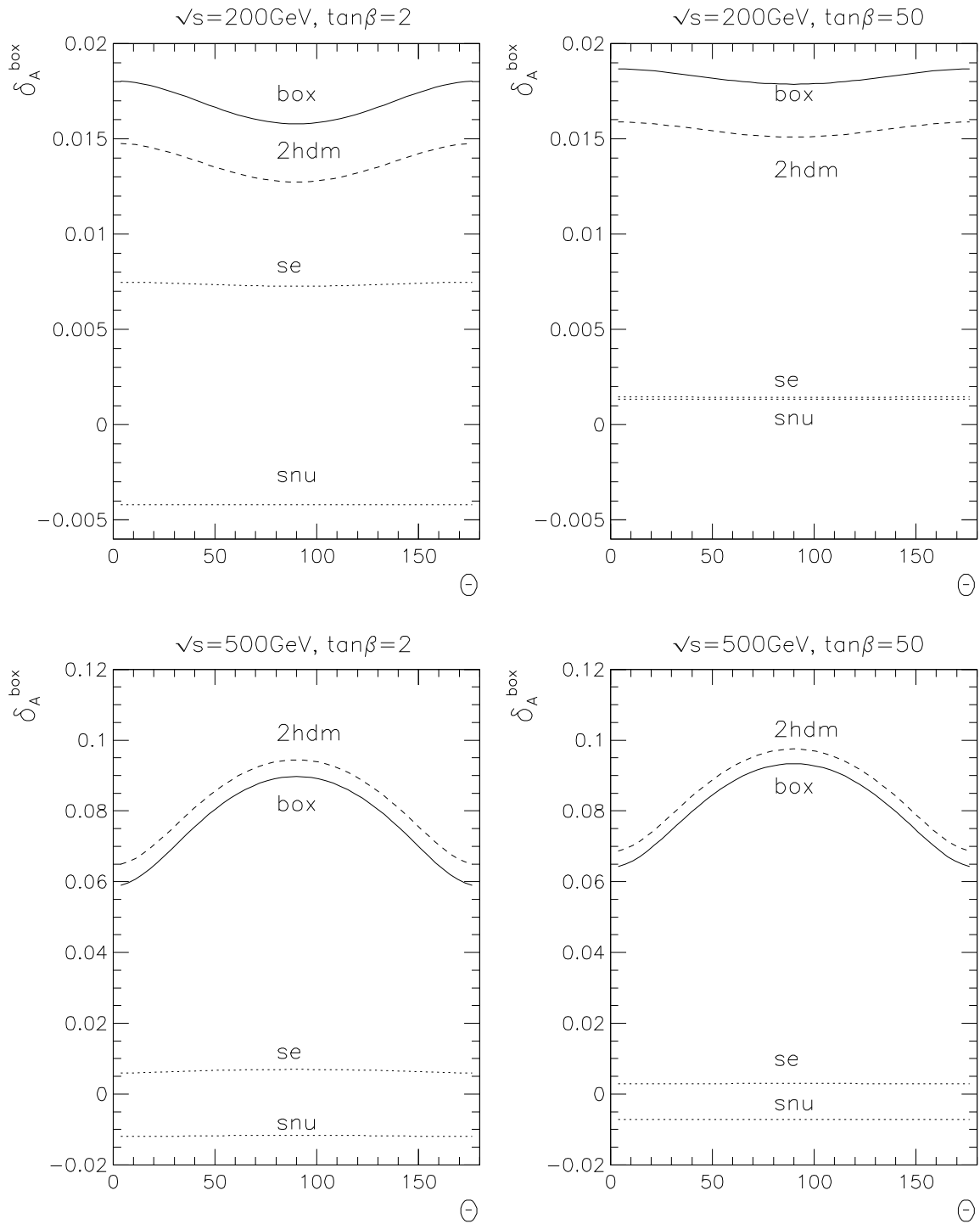
\item{\bf Fig.7}  Relative box contribution $\delta^{box}_A$
  for  $ e^+ e^- \rightarrow A^0 h^0 $
  at $\sqrt{s}=200$ GeV, 500 GeV and $\tan\beta=2$, 50.
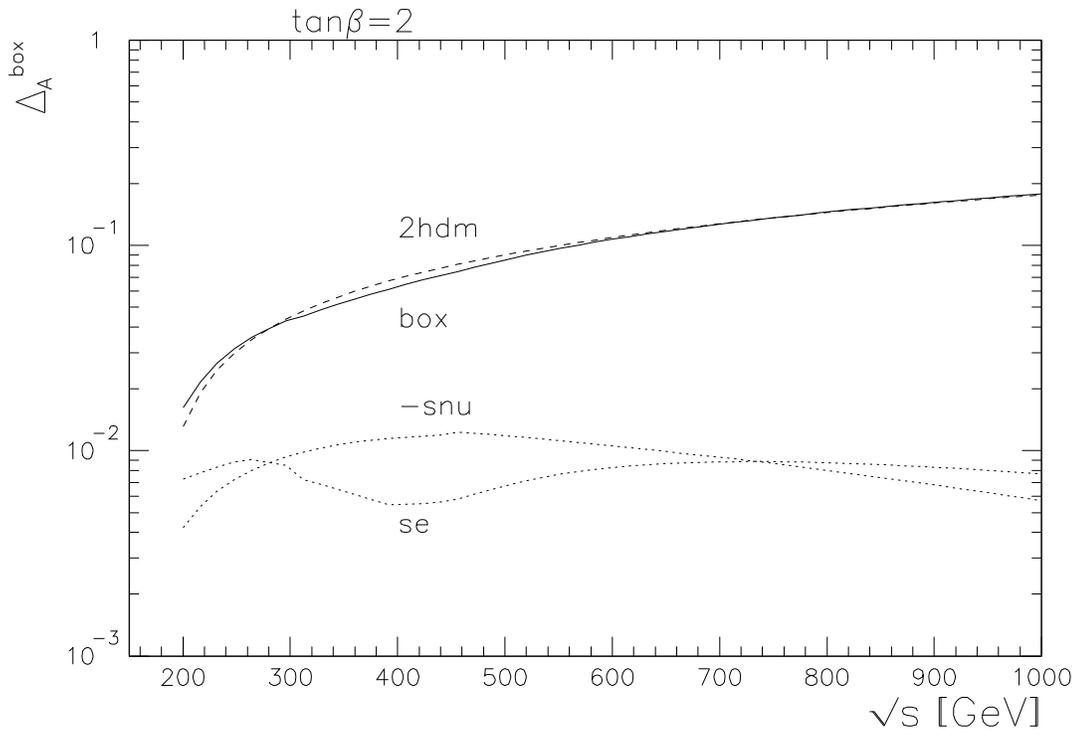
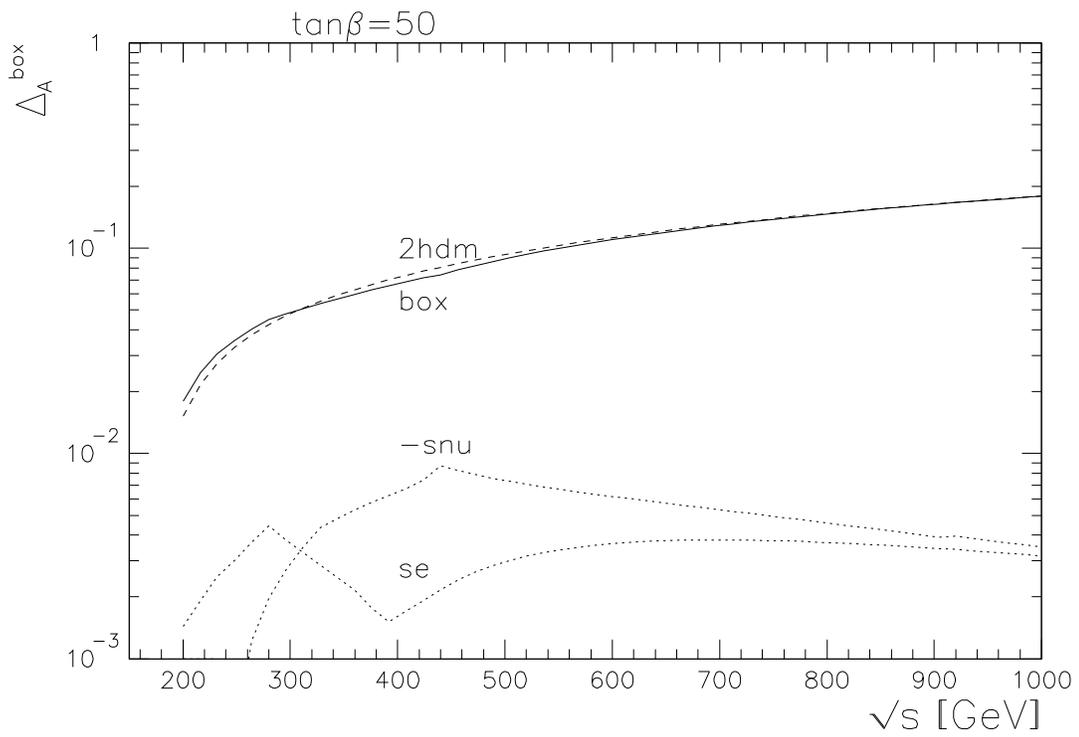
\item{\bf Fig.8}  Relative box contribution $\Delta^{box}_A$
  for $M=\mu=150$ GeV.
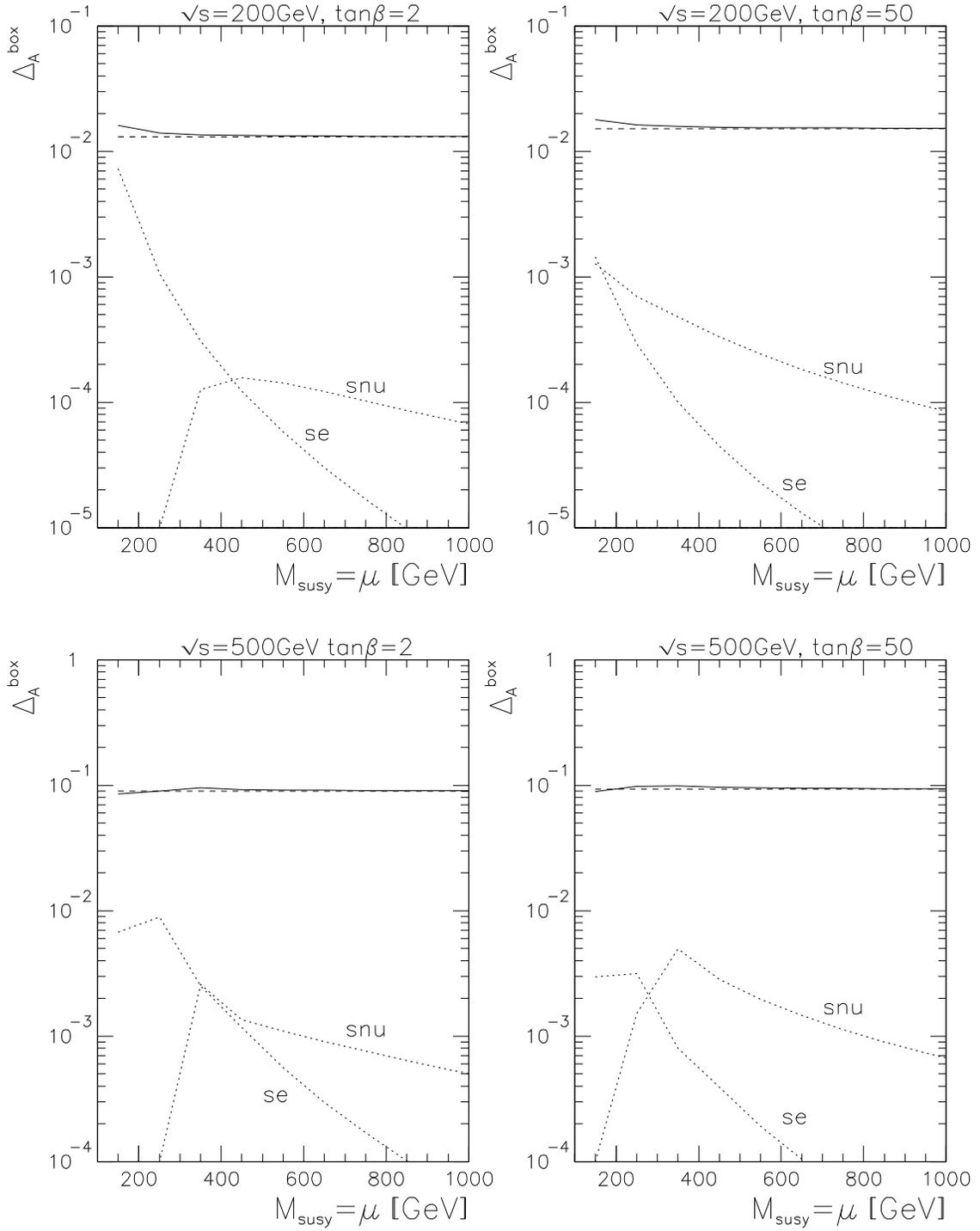
\item{\bf Fig.9}   $\Delta^{box}_A$
  at $\sqrt{s}=200$ GeV, 500 GeV and $\tan\beta=2$, 50.
\end{itemize}

\end{document}